# Video coding technique
# with parametric modeling of noise

*Olgierd Stankiewicz*

*Chair of Multimedia Telecommunications and Microelectronics, Poznan University of Technology,*

*Pl. Marii Skłodowskiej-Curie 5, 60-965 Poznań, Poland*

*olgierd.stankiewicz@put.poznan.pl*

**Abstract**

This paper presents a video encoding method in which noise is encoded using a novel parametric model representing spectral envelope and spatial distribution of energy. The proposed method has been experimentally assessed using video test sequences in a practical setup consisting of a simple, real-time noise reduction technique and High Efficiency Video Codec (HEVC). The attained results show that the use of the proposed parametric modeling of noise can improve the subjective quality of reconstructed video by approximately 1.8 Mean Opinion Scope (MOS) points (in 11-point scale) related to the classical video coding. Moreover, the present work confirms results attained in the previous works that the usage of even sole noise reduction prior to the encoding provides quality increase.

## 1. Introduction

In recent years, the overall share of video transmission in total internet traffic was increasing systematically. Some sources estimate [1] that video broadcasters such as YouTube, Netflix, and the VoD (Video on Demand) services already attract approximately one-third of all internet users. It was estimated [2] that video accounted for 73% of total internet traffic in 2016 and it will likely increase to 82% by 2021. A vast number of subscribers, approximately 125 million [3], force service providers to search for new, highly efficient technical solutions. This need stimulates both the development of internet infrastructure and development of more efficient video compression technologies.

During three decades of research on video coding, many techniques have been developed in academia, research institutes and industry [4]. The most influential incubators of successive generations of video compression techniques are Motion Picture Experts Group (MPEG) working on behalf of International Organization for Standardization (ISO) and International Electrotechnical Commission (IEC), and VCEG (Video Coding Expert Group) who work on behalf of International Telecommunication Union (ITU). The cooperation of these two bodies, e.g. under the form of Joint Collaborative Team on Video Coding (JCT-VC), resulted in the following development of technologies adopted as international standards: MPEG-2 [5], Advanced Video Coding (AVC) [6][7], High Efficiency Video Coding (HEVC) [8][9]. These standards correspond to successive generations of compression technology, each of which provides approximately 50% reduction of bitrate required for the same quality of decoded video with respect to the previous one (e.g. HEVC with respect to AVC) [4][7]. Such gain is achieved through the use of increasingly complex compression algorithms. Currently, MPEG and VCEG are jointly working on Versatile Video Coding (VVC), which is expected to be standardized in 2020 [10]. Additionally, industrial consortia have also proposed several complete compression technologies (e.g. recent AV1 codec [11]).

Natural video content, captured with real cameras, is inherently characterized by the presence of noise originating from photosensitive electrical circuits, often referred to as film grain. Although the quality of optical sensors improves over time, which theoretically allows for higher Signal-to-Noise Ratios (SNRs), the same occurs in the context of camera resolutions and frame rates. Higher resolutions lead to smaller pixels on sensors, while higher frame rates enforce shorter exposure times. As a result of both of these factors, the amount of noise present in video often remains significant, although the exact amount and characteristics of noise varies [12]. Even with modern camera technology, noise remains noticeable, especially under poor illumination or for low-end consumer cameras [13].

Noise in video disturbs many processes of video compression, such as inter prediction, intra prediction, mode selection, and quantization. Since noise is random by nature, it cannot be efficiently predicted, and its existence reduces compression performance. In practice, this implies that an encoder would spend bits for an inefficient representation of noise. Therefore, a common solution involves removal or reduction of noise prior to encoding (see [14], [15], [16]). Although the advantage of such an approach is that the bitrate is significantly reduced, information regarding noise is not encoded at all. This represents a disadvantage, since noise is perceptually important to viewers [17] as the absence of noise causes a sense of unnaturalness.

This paper addresses the mentioned aforementioned issue of compression of video with noise. Instead of simple noise removal, the paper proposes to estimate the noise parameters in the encoder and transmit these parameters in the encoded bitstream. This is a similar approach to the one already successfully used in audio coding, e.g. in the Perceptual Noise Substitution (PNS) tool in Advanced Audio Codec [18]. A similar idea in the context of video has been employed in previous works [19], [20] and [21], where authors also proposed the use of parametric noise model and noise synthesis on the decoder side. Notably, works [19] and [20] were performed in the context of older AVC video coding technology. These works provided very limited experimental verification since only a few sequences were used. Moreover, in [20], only objective results are provided, while [19] provided only very limited subjective quality results. Work [21] is performed in the context of the AV1 video codec and it also considers parametric noise modeling. In particular, a two-dimensional autoregressive model is applied for modeling of the spectrum and the amplitude of the synthetized noise is modeled with piecewise linear function. Also in this work a very limited set of experimental results is provided (based on four sequences) with no formal subjective evaluation performed. Notably, none of the aforementioned works assess whether the attained gains originate solely from noise reduction or from the addition of synthetic noise. These issues are considered in the present paper.

Therefore, the significant contributions of this paper include:

- Proposal of a novel parametric model of noise, consisting of two components: Spatial Distribution of Energy (low-resolution video) and Spectral Envelope (encoded with Log-Area-Ratio coefficients [22]).
- Implementation of the abovementioned idea on top of the state-of-the-art HEVC video coding technology.
- Exhaustive experimental results of subjective evaluation, providing a comparison of the performance of the proposed method versus classical video coding scheme - with and without noise reduction.

## 2. Proposed coding scheme

The general idea of the proposal (Fig. 1) consists in coding of video with the use of two layers:

- The base layer, which contains useful content that can be efficiently modeled with existing predictive tools.
- The noise layer, which contains unpredictable random content that can be modeled statistically.

While the base layer is coded using standard methods, only a parametric model is transmitted for the noise layer. At the decoder, the parametric model is used to synthesize noise, which resembles the original one, inputted to the encoder.

## 2.1. Encoder

The first step of the processing on the encoder side (Fig. 1) is decomposition of the input video $I(x, y)$ into two layers: the base layer video $V(x, y)$ and the noise layer $N(x, y)$. The noise layer video $V(x, y)$ is attained with the use of noise reduction. Although the proposed idea is general and disregards any particular technique, the use of noise reduction with different quality may obviously influence the attained results. The noise layer video $N(x, y)$ is calculated as a difference between the input video $I(x, y)$ and the base layer video $V(x, y)$.

The base layer video $V(x, y)$ is encoded with a general-purpose video encoder (Fig. 1), such as AVC, HEVC or AV1. Due to the decomposition, the base layer contains significantly reduced amount of noise and can thus be efficiently modeled using predictive compression tools. In particular, the potential benefits of such an approach include:

- Intra and inter prediction may be more efficient, thus making the prediction error potentially smaller,
- Motion vectors for inter-frame prediction may be more consistent and represent the real motion in video, thus being more susceptible to compression.
- The structure of the compressed image and selection of encoding modes may be more regular. Therefore, control elements (e.g. signaling of CU hierarchy partitioning in HEVC) may be compressed more efficiently in the course of entropy encoding.

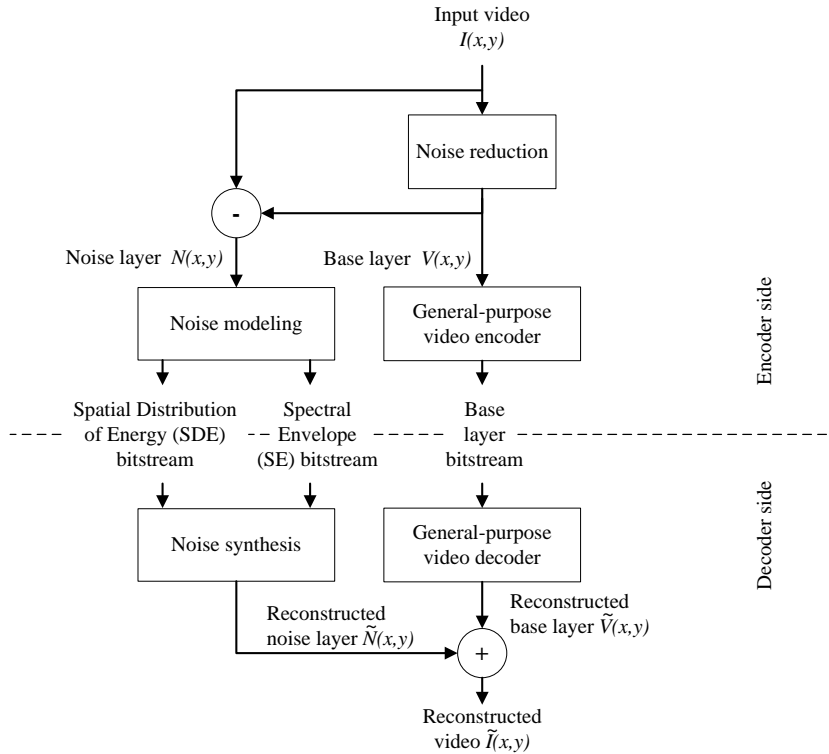

Fig. 1. Scheme of the proposed video coding using parametric modeling of the noise.

The noise layer $N(x, y)$, on the other hand, contains unpredictable random content. Encoding of such signals with general-purpose video coding techniques is very inefficient. Therefore, the present paper proposes parametric modeling the noise layer. For this purpose, two components that describe spatial and spectral characteristics of the modeled noise are proposed:

- Spatial Distribution of Energy (SDE), and
- Spectral Envelope (SE).

Whole noise modeling is performed independently for each color channel (e.g. Y, $C_b$, $C_r$). As such, for the sake of brevity, the following explanation is provided in the context of a single channel.

Spatial Distribution of Energy (SDE) conveys information regarding the amount of noise residing in particular spatial regions of the noise layer video. This is represented with small size video, the values of which correspond to the amount of energy of noise. SDE is estimated as follows.

Firstly, the noise layer video is divided into square non-overlapping blocks. In each of those blocks, noise energy is measured with the use of root mean square (RMS) metric. RMS values associated with respective blocks constitute a map of Spatial Distribution of Energy. Noise energy distribution can change over time; therefore SDE is calculated for each frame of the noise layer video individually using the following formula (1):

$$SDE(m,n) = \sqrt{\sum_{x=m\cdot\beta}^{m\cdot\beta+\beta-1} \sum_{y=n\cdot\beta}^{n\cdot\beta+\beta-1} |N(x,y)|^2} \quad , \tag{1}$$

where:

$N(x,y)$ – value in the noise layer video at pixel coordinates $(x,y)$ in the current frame,

$(x,y) \in \{0..W-1\} \times \{0..H-1\}$,

$W, H$ – width and height (in pixels) of the processed video,

$\beta$ – block size used (e.g. 30 in experiments),

$SDE(m,n)$ – value of the Spatial Distribution of Energy at coordinates $(m,n)$,

$(m,n) \in \{0..W_\beta-1\} \times \{0..H_\beta-1\}$,

$W_\beta = \left\lfloor \frac{W}{\beta} \right\rfloor, H_\beta = \left\lfloor \frac{H}{\beta} \right\rfloor$ – width and height of the Spatial Distribution of Energy image,

$\lfloor \cdot \rfloor$ – floor rounding operation.

Notably, the resolution of SDE ($W_\beta \times H_\beta$) is smaller than the resolution of the input video ($W \times H$) depending on the size of blocks $\beta$ used. Therefore the bitstream required to represent SDE is negligible (about 1% depending on the case - see: Table 7) in relation to the bitstream required to represent the video

Spectral Envelope (SE) conveys information regarding the spectrum of the signal in the noise layer. It is represented by the transfer functions of Infinite Impulse Response (IIR) filters in horizontal and in vertical direction. SE is estimated with the use of a linear predictive model. The proposed technique is very similar to Linear Predictive Coding (LPC) known from audio compression [22]. First, the values in the noise layer video $N(x,y)$ are normalized with respect to RMS values that were calculated beforehand and stored in SDE (2):

$$N_N(x,y) = \frac{N(x,y)}{SDE(\lfloor m/\beta \rfloor, \lfloor n/\beta \rfloor)}. \tag{2}$$

This step is required in order to attain stability of the LPC algorithm, which would be otherwise unstable due to energy fluctuations in the signal [22].

Next, the linear predictive model is applied in the horizontal and vertical direction in order to estimate IIR filters parameters. This is attained by the construction of one-dimensional signals of concatenated rows and columns of $N_N(x,y)$, respectively:

$$N_{N\,horizontal}(x) = [\, N_N(\cdot,0)\ N_N(\cdot,1)\,...\,N_N(\cdot,H-1)\,]\,, \qquad N_{N\,vertical}(y) = \begin{bmatrix} N_N(0,\cdot) \\ N_N(1,\cdot) \\ ... \\ N_N(W-1,\cdot) \end{bmatrix}. \tag{3}$$

Therefore, normalized noise layer video $N_N(x,y)$ is rearranged into a single row signal $N_{N\,horizontal}(x)$ and into a single column signal $N_{N\,vertical}(y)$. For each of these, the classical linear predictive model of order $p$ is used, which can be expressed by the general approximation equation (4) for a certain signal $S(k)$:

$$S(k) \approx a_1 \cdot S(k-1) + \ldots + a_p \cdot S(k-p) \,. \tag{4}$$

The linear prediction weights $a_1 \ldots a_p$ are coefficients of an IIR filter (6) that represent the spectrum of the analyzed signal $S(k)$.

$$H(z) = \frac{1}{1 - \sum_{k=1}^{p} a_k \cdot z^{-k}} \,. \tag{5}$$

Linear prediction is performed for both the horizontal and vertical direction, so that $S(k) \leftarrow N_{N\,horizontal}(x)$ and $S(k) \leftarrow N_{N\,vertical}(y)$, which yields IIR filter transfer functions $H_{horizontal}(z)$ and $H_{vertical}(z)$. This is performed individually for each frame of the noise layer video.

The proposed statistical model of noise can be summarized to consist of:

- An image of the Spatial Distribution of Energy $SDE(m, n)$,

- Spectral Envelope (SE) expressed by the transfer functions of filters $H_{horizontal}(z)$ and $H_{vertical}(z)$ ,

both for each frame of the noise layer. The exemplary amounts of data required for transmission of the proposed parametric model of the noise are presented in Table 7 (in experimental results section), e.g. for HD sequences (class B1) it is required to transmit about 4 kbit/s of SE data and 4 kbit/s for SDE data.

## 2.2. Decoder

On the decoder side (Fig. 1), the noise layer is synthesized using the coded parameters of the noise (Fig. 2).

First, white Gaussian noise video $\widetilde{N}_{WG}(x, y)$ is generated. Resolution of this video is the same as that of the input video ($W \times H$). Next, the generated noise is filtered in the horizontal and in the vertical direction using IIR filters with transfer functions $H_{horizontal}(z)$ and $H_{vertical}(z)$, respectively. The result is noise video $\widetilde{N}_{SS}(x, y)$, which is spectrally shaped to resemble the spectral properties of the original noise layer $N(x, y)$. Next, the values from $\widetilde{N}_{SS}(x, y)$ are scaled (multiplied) with the corresponding values from the Spatial Distribution of Energy $\widehat{SDE}(m, n)$, which is also reconstructed at the decoder side. This provides the reconstructed noise layer $\widetilde{N}(x, y)$, which resembles both the spectral and spatial properties of the original noise layer $N(x, y)$.

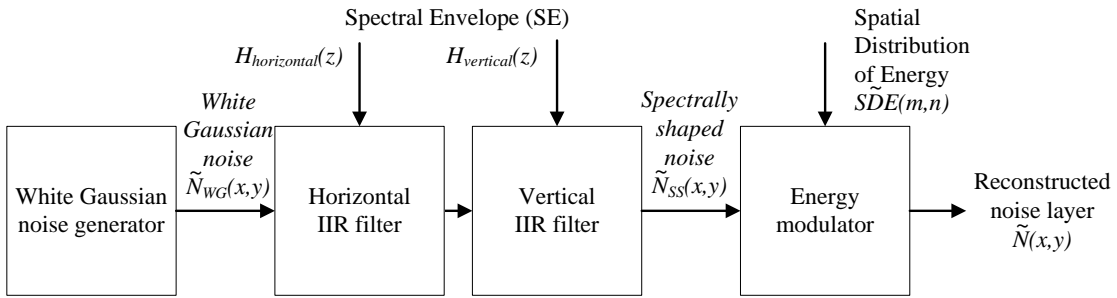

Fig. 2. Noise synthesis scheme in the proposed decoder.

Finally, the synthesized noise layer $\widetilde{N}(x, y)$ is added (Fig.1) to the reconstructed base layer $\widetilde{V}(x, y)$ to produce the output reconstructed video $\tilde{I}(x, y)$.

## 3. Implementation

For the sake of experimental verification, the general idea of the proposal described in Section II is further elaborated and implemented as described below. The whole processing was performed in $YC_bC_r$ color space.

Noise reduction (Fig. 1) was realized with the use of the motion-compensated technique implemented in "mv-tools" motion compensation software package [23], which is a plug-in for the VirtualDub/AviSynth video scripting framework [24]. It consists of finite impulse response (FIR) filtering along the direction of movement in the video, with weights adaptively selected depending on the content.

The input frames neighboring in time (next and previous ones) are used to reduce noise in the currently processed frame, denoted as $Frame_f$ (Fig. 3). At first, each neighboring frame, denoted as $Frame_n$, $n \in \{f - K, ..., f - 2, f - 1, f + 1, f + 2, ..., f + K\}$ is used to predict the current frame $Frame_f$ with the use of block-based motion compensation, thereby yielding $Prediction_n$. This step is similar to the motion vector search known from video coding. Next, the predicted images $Prediction_n$ are averaged together with the current frame $Frame_f$, with weights depending on the similarity of particular predicted image $Prediction_n$ to the current frame $Frame_f$. A variant of the technique with six neighboring frames (3 next and 3 previous ones, thus $K = 3$) was used, due to its good performance versus computational complexity (Fig. 3). More details can be found in [23].

For the role of General-Purpose Video Encoder and Decoder (Fig. 1), HEVC video coding technology was selected along with its implementation in MPEG HEVC model software HM 13.0. The coding configuration recommended by JCT-VC in *"Common Test Conditions"* [25] was used, e.g. Intra period of 32 frames.

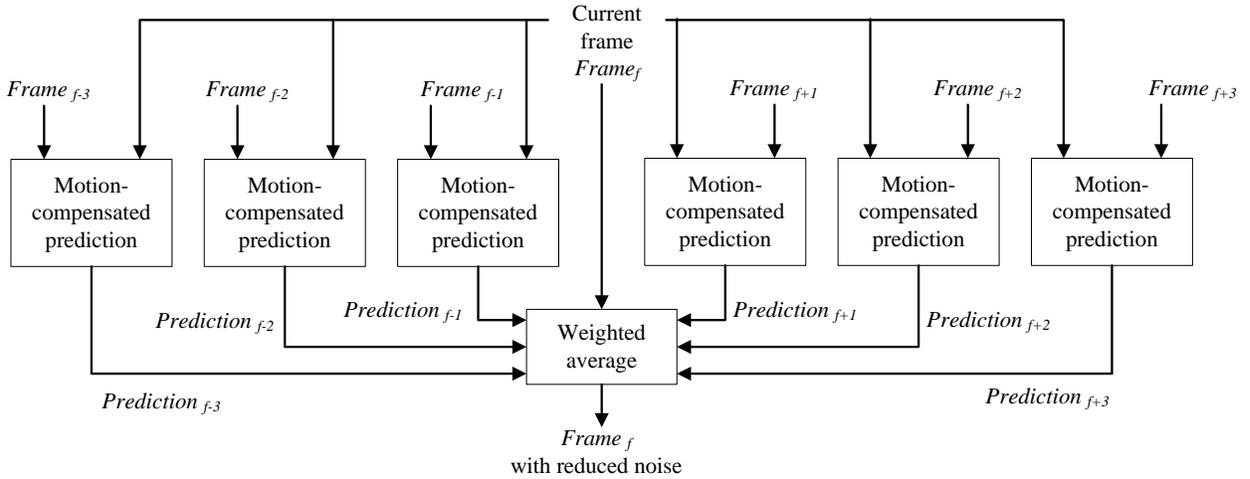

Fig. 3. Scheme of the utilized noise reduction technique [23].

The Spectral Distribution of Energy (SDE) was calculated in square blocks of $\beta = 30$ in size. This yielded significantly smaller SDE images than the original input resolution. For example, an HD input image (1920×1080) the size of SDE image was 64×36.

For the compression of SDE, HEVC was also selected. Uniformly (for all bitrates and sequences), QP value of 30 was used. Notably, such a small resolution of image (as 64×36 pixels) fits into a single HEVC Coding Unit (CU), thus generating a very small compressed bitstream.

Spectral Envelope (SE) was estimated using the linear prediction of order $p = 10$. IIR filter parameters were estimated with the use of the Levinson-Durbin recursion algorithm [26]. For the sake of compression, PARCOR $r_k$ coefficients were used instead of classical filter coefficients $a_k$. $r_k$ coefficients, also called reflection coefficients, which are the by-product of the Levinson-Durbin algorithm are known to be much less vulnerable for lossy compression than $a_k$ coefficients. A known solution based on log-area-ratio (LAR) [22] representation was used (6):

$$R_k = log \frac{1 - r_k}{1 + r_k}. \tag{6}$$

All of the LAR coefficients $R_k$ were coded using 8-bit representation.

### 4. Experimental results

The main idea of the proposal consists in encoding noise parameters, and using them in the decoder to synthesize noise, which resembles but is not identical to that on the input. Therefore, the estimation of such popular objective quality parameters such as Mean Square Error (MSE), Peak Signal-to-Noise Ratio (PSNR), or Structured Similarity Index (SSIM) is irrelevant [27]. Thus, the methodology used during the present experiments was based on subjective quality assessment. In particular, it was based on tests performed for the evaluation of results of *"Call for Proposals"* [28] issued by MPEG in search for video compression tools for HEVC. In particular, the same viewing method (DSIS - Double Stimulus Impairment Scale) [30], the same sequences (Table 1), and the same bitrates (Table 2) [29] were used in this paper. Please note that an additional bitrate Rate 0 (lower than Rate 1) has been included in addition to [28] in order to allow better HEVC-based codec comparison. In total, experiments used 19 sequences divided into 6 classes with various resolutions and frame-rates.

Table 1. Sequences used in the experimental evaluation [28][29].

| Sequence class | Resolution | Sequence | Frame rate |
|---|---|---|---|
| Class A | 2560×1600 | S1.  Traffic | 30 |
| | | S2.  PeopleOnStreet | 30 |
| | | S3.  SteamLocomotive | 60 |
| Class B1 | 1920×1080 | S4.  Kimono | 24 |
| | | S5.  ParkScene | 24 |
| Class B2 | 1920×1080 | S6.  Cactus | 50 |
| | | S7.  BQTerrace | 60 |
| | | S8.  Basketball Drive | 50 |
| Class C | 832×480 | S9.  RaceHorses | 30 |
| | | S10. BQMall | 60 |
| | | S11. PartyScene | 50 |
| | | S12. BasketballDrill | 50 |
| Class D | 416×240 | S13. RaceHorsesLow | 30 |
| | | S14. BQSquare | 60 |
| | | S15. BlowingBubbles | 50 |
| | | S16. BasketballPass | 50 |
| Class E | 1280×720 | S17. FourPeople | 60 |
| | | S18. Johnny | 60 |
| | | S19. KristenAndSara | 60 |

During the assessment performed for the sake of the present work the Double Stimulus Impairment Scale (DSIS) method in accordance with ITU Recommendation BT.500 [30] was used. Viewers interchangeably viewed the reference (uncompressed original video) and tested case (video encoded and decoded with the tested codec), which was repeated twice for each case before the voting. As per the evaluation performed in work [29], a particular variation Mean Opinion Score (MOS) was applied, using quality scale from 0 to 10, instead of the classical "impairment" quality scale.

Table 2. Bitrates selected for experimental evaluation for classes of sequences [kbit/s] [28][29]. Additional "Rate 0" has been added in addition to [28].

| Sequence class | Rate 0 | Rate 1 | Rate 2 | Rate 3 | Rate 4 | Rate 5 |
|---|---|---|---|---|---|---|
| Class A | 1600 | 2500 | 3500 | 5000 | 8000 | 14000 |
| Class B1 | 650 | 1000 | 1600 | 2500 | 4000 | 6000 |
| Class B2 | 1300 | 2000 | 3000 | 4500 | 7000 | 10000 |
| Class C | 256 | 384 | 512 | 768 | 1200 | 2000 |
| Class D | 192 | 256 | 384 | 512 | 850 | 1500 |
| Class E | 192 | 256 | 384 | 512 | 850 | 1500 |

The tested codecs were configured to fit the encoded sequences within the required bitrates (Table 2), which were identical for all codecs. As such it was possible to directly compare the quality since the compared cases shared the same bitrate. The compared codecs were as follows:

A. Anchor – original video (with noise) encoded with standard HEVC.

B. Noise reduction without synthesis – video is denoised prior to the coding but the noise is not synthesized at the decoder.

C. Proposal –video encoded according to the proposal (as base and noise layer) including the addition of synthetic noise to the reconstructed video at the decoder side.

The idea behind including case B was to determine whether gains of the proposal originated from the denoising of the input sequence or from the addition of the synthetic noise in the decoder. In fact, since the bitstream overhead of the proposed noise parameters is negligible, it can be noted that case B is equivalent to the proposal (C) but without adding of the synthetic noise in the decoder. Although there is potential gain (of about 1% depending on the case - see: Table 7) from not sending noise parameters, it would not allow to change the quality of reconstructed video, e.g. to increase quality of the base video layer by selecting smaller quantization parameter (QP) (Table 3).

Table 3. Quantization parameter (QP) for the HEVC video encoder case A (anchor) and QP value corrections ΔQP for the tested cases B and C (proposed).

| Sequence | QP values for the case A (anchor) | | | | | | ΔQP of values for cases: B and C (proposed) | | | | | |
|---|---|---|---|---|---|---|---|---|---|---|---|---|
| | Rate 0 | Rate 1 | Rate 2 | Rate 3 | Rate 4 | Rate 5 | Rate 0 | Rate 1 | Rate 2 | Rate 3 | Rate 4 | Rate 5 |
| S1.  Traffic | 36 | 33 | 30 | 28 | 25 | 22 | 0 | -1 | -1 | -2 | -2 | -3 |
| S2.  PeopleOnStreet | 46 | 43 | 40 | 37 | 33 | 28 | 0 | 0 | 0 | -1 | -1 | -1 |
| S3.  SteamLocomotive | 36 | 33 | 31 | 29 | 27 | 24 | 0 | -1 | -2 | -2 | -4 | -4 |
| S4.  Kimono | 36 | 33 | 30 | 26 | 23 | 21 | -1 | -1 | -2 | -1 | -2 | -3 |
| S5.  ParkScene | 38 | 35 | 32 | 29 | 26 | 24 | -1 | -1 | -1 | -1 | -2 | -3 |
| S6.  Cactus | 39 | 35 | 32 | 29 | 27 | 25 | -1 | 0 | -1 | -1 | -2 | -3 |
| S7.  BQTerrace | 37 | 35 | 32 | 30 | 28 | 27 | 0 | -2 | -2 | -2 | -3 | -4 |
| S8.  Basketball Drive | 39 | 35 | 32 | 29 | 27 | 25 | -1 | 0 | -1 | -1 | -2 | -3 |
| S9.  RaceHorses | 41 | 39 | 37 | 34 | 31 | 28 | 0 | -1 | -1 | -1 | -1 | -2 |
| S10. BQMall | 43 | 40 | 37 | 34 | 31 | 27 | -1 | -1 | 0 | -1 | -1 | -1 |
| S11. PartyScene | 44 | 42 | 40 | 38 | 34 | 31 | 0 | -1 | 0 | -1 | 0 | -1 |
| S12. BasketballDrill | 42 | 39 | 37 | 33 | 30 | 27 | 0 | 0 | -1 | 0 | -1 | -2 |
| S13. RaceHorsesLow | 35 | 33 | 30 | 28 | 25 | 21 | 0 | -1 | -1 | -1 | -2 | -3 |
| S14. BQSquare | 37 | 35 | 32 | 30 | 27 | 23 | -1 | -1 | -2 | -2 | -2 | -2 |
| S15. BlowingBubbles | 37 | 35 | 32 | 31 | 27 | 23 | 0 | -1 | 0 | -1 | -1 | -2 |
| S16. BasketballPass | 38 | 35 | 32 | 30 | 27 | 23 | -1 | 0 | 0 | 0 | -1 | -2 |
| S17. FourPeople | 44 | 41 | 38 | 35 | 31 | 26 | 0 | 0 | 0 | 0 | -1 | -1 |
| S18. Johnny | 38 | 35 | 32 | 30 | 26 | 23 | 0 | 0 | -1 | -1 | -1 | -3 |
| S19. KristenAndSara | 41 | 38 | 35 | 32 | 28 | 25 | 0 | 0 | -1 | 0 | -1 | -3 |

The subjective evaluation tests were performed in many, 20-minute sessions, performed in different time, in order not to overstress the visual systems of viewers. In total, there were 94 subjects. Due to the high number of subjects the attained 95% confidence intervals are very small - to of order of ±(0.06÷0.18). Confidence intervals have been marked on all graphs presenting the results (Figs. 4-9).

The example frames of the sequences are presented in Fig. 10. For the sake of comparison, results for the middle bitrate (R3) are shown. Notably, the noise synthesized at the decoder with the proposed method (Fig. 10e) is practically indistinguishable from the original noise (Fig. 10a). Moreover it can be seen that Anchor encoding (Fig. 10c) was not able to represent the original noise in the compressed bitstream and also wasted some bits trying to do so, which, due to limited total bitrate available, deteriorated the quality of the video itself.

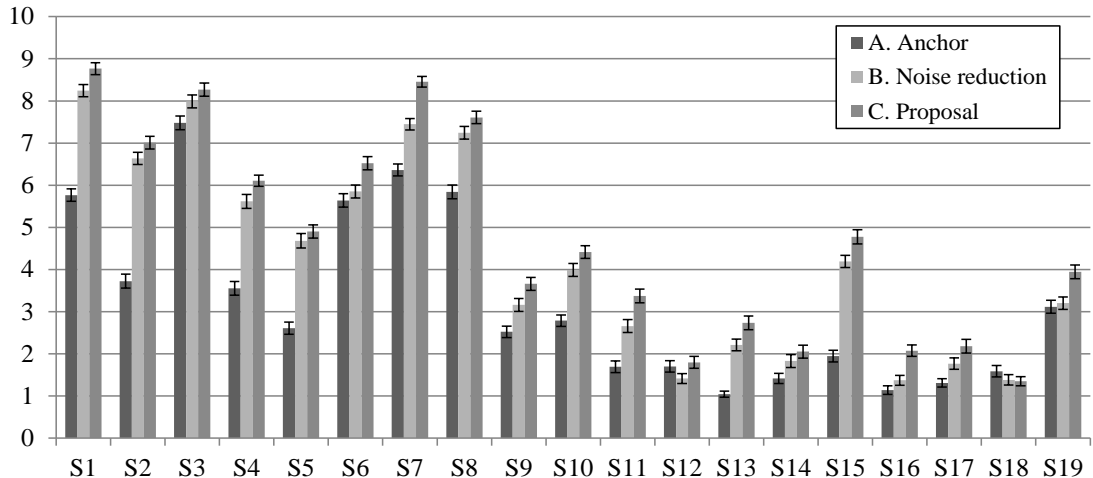

Fig. 4. Results of subjective evaluation for Rate 0, with marked 95% confidence intervals.

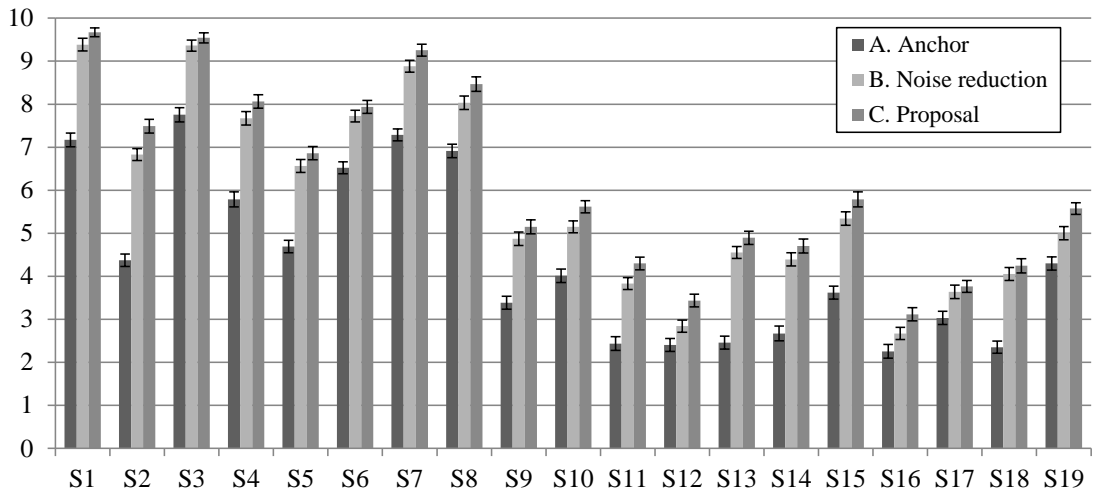

Fig. 5. Results of subjective evaluation for Rate 1, with marked 95% confidence intervals.

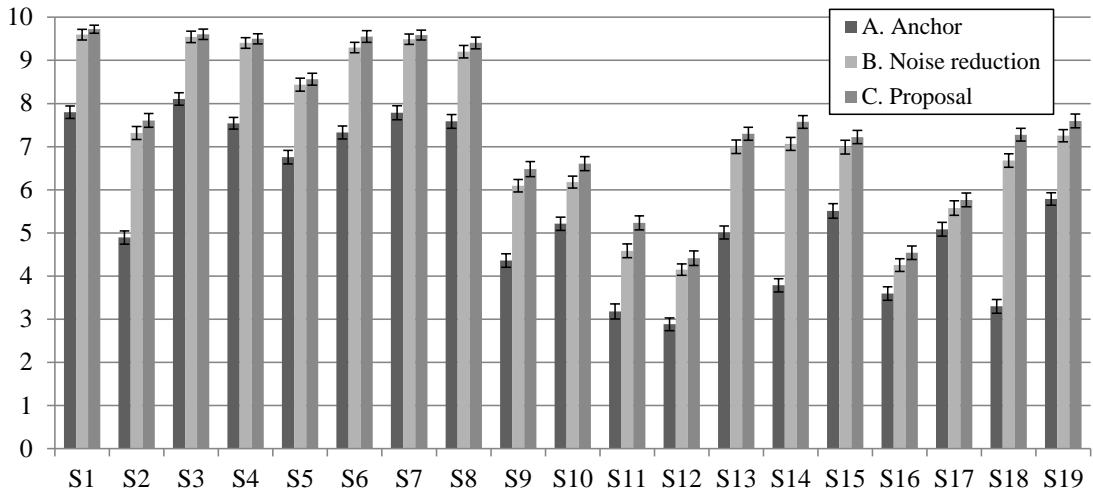

Fig. 6. Results of subjective evaluation for Rate 2, with marked 95% confidence intervals.

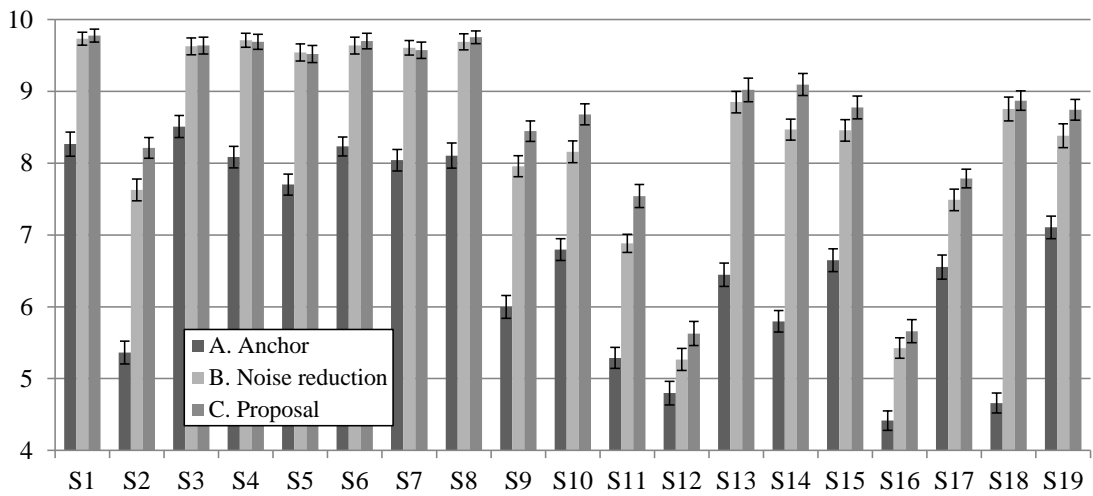

Fig. 7. Results of subjective evaluation for Rate 3, with marked 95% confidence intervals. Please note the displayed MOS scale range.

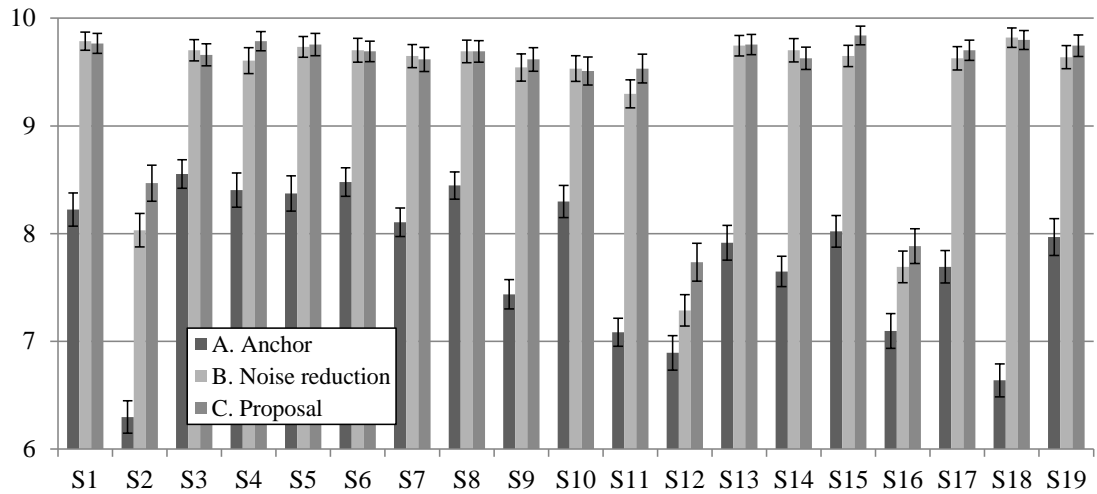

Fig. 8. Results of subjective evaluation for Rate 4, with marked 95% confidence intervals. Please note the displayed MOS scale range.

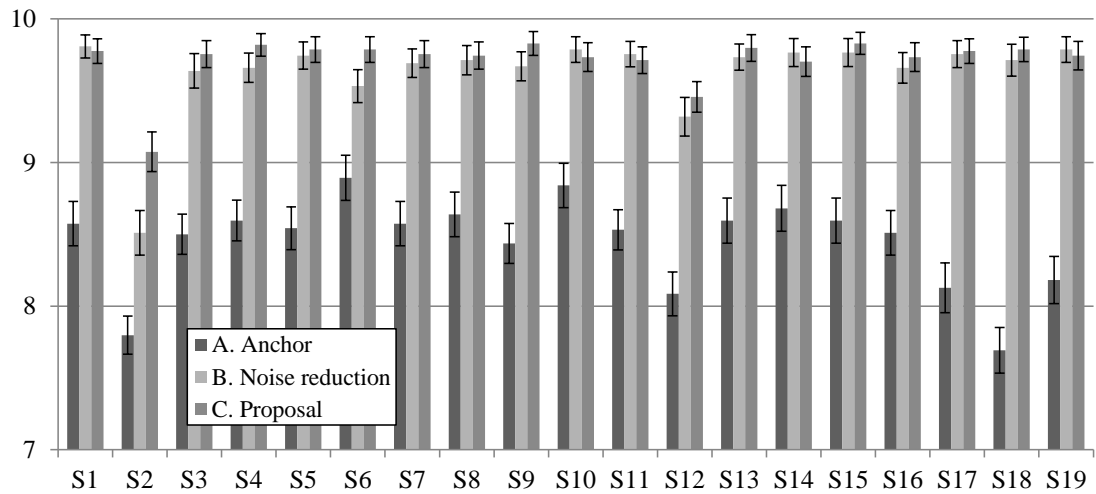

Fig. 9. Results of subjective evaluation for Rate 5, with marked 95% confidence intervals. Please note the displayed MOS scale range.

Based on the results it is evident that - in practically all of cases - the usage of noise reduction prior to compression (B) statistically significantly increases the quality of compressed video versus the anchors (A). Except of the negligible cases of sequences S12 and S18 for a single rate-point Rate 0 (Fig. 4), none of the confidence intervals for (B) and (A) overlap. The observed gain is approximately 0.5 to 2.0 MOS points. This confirms the results attained in previous works [15] and [16]. It can be noticed that the highest bitrate (R5) has been selected too high, since the scores provided by the subjects were saturated at about 10 (maximum), making the observation of differences between the analyzed cases more difficult. This rate has been used in the work [28] in order to allow comparison of AVC versus HEVC and is impractical in HEVC versus HEVC comparisons.

Moreover, the usage of the proposed parametric modeling of noise (C) increases quality gain even further, to range of about 1.0 to 2.0 MOS points. This difference is statistically significant in almost all cases with relation to the anchors (A), again with exception of sequences S12 and S18 for Rate 0. In relation to the noise reduction (B), the increase of quality of the proposed case (C) is about 0.2 to 0.5 MOS. In most of the cases, the quality gain is statistically significant, especially for lower bitrates. For example, the proposal (C) is statistically    significantly better than (B) in the case of 15 sequences out of 19 sequences in total, when Rate 0 is considered (Fig. 4) and the case of 16 sequences out of 19 sequences in total, when Rate 1 is considered (Fig. 5).

This results is slightly worst when higher bitrates are considered – e.g. when Rate 4 is considered, the proposal (C) is better than (B) in the case of only 4 sequences. In other cases, the confidence intervals overlap, but also it can be noted, that the proposed usage of the parametric modeling of noise (C) is never worse than (B) or (A).

The average results of the evaluation are provided in Tables 4 and 5. Of course, averaging is not a statistically correct method, e.g. it disallows comparison of the confidence intervals. Nevertheless, the respective results have been provided to indicate trends in the analyzed data. In particular, it can be seen that on average the gain of the noise reduction (B) versus the anchor (A) is about 1.44 MOS points. The proposed parametric modeling of noise (C) is about 1.7 MOS points better than the anchor (A) on average. Moreover, gains attained through the use of the proposed method are approximately the same for all of sequences, classes, and rate-points. It can also be concluded that the use of the proposed parametric modeling of noise undoubtedly provided increased quality in relation to noise reduction only, with the attained gain of approximately 0.26 MOS points. This value is lower than the quality gain attained with the use of noise reduction (approximately 1.44 MOS points), but it is entirely justified if computational costs are taken into consideration. Table 6 presents the computational complexity results based on execution time for particular modules. The test was performed using an Intel Core i7-3770K processor working at 3.5GHz. While the use of noise reduction increased the working time of the codec by approximately 1% - the parametric modeling of noise did not practically change the working time at all.

Table 4. Subjective (MOS) results for different bitrates averaged overs sequences.

| Rate | A. Anchor | B. Noise reduction | C. Proposal | B gain = B - A | C gain = C - A |
|------|-----------|--------------------|-------------|----------------|----------------|
| Rate 0 | 3.22 | 4.26 | 4.74 | 1.04 | 1.52 |
| Rate 1 | 4.39 | 5.83 | 6.20 | 1.44 | 1.81 |
| Rate 2 | 5.55 | 7.27 | 7.56 | 1.72 | 2.01 |
| Rate 3 | 6.68 | 8.38 | 8.64 | 1.7 | 1.96 |
| Rate 4 | 7.77 | 9.34 | 9.43 | 1.57 | 1.66 |
| Rate 5 | 8.44 | 9.63 | 9.72 | 1.19 | 1.28 |
| **Average** | **6.01** | **7.45** | **7.72** | **1.44** | **1.71** |

Table 5. Subjective (MOS) results for different sequences and classes averaged over bitrates.

| Class | Sequence | A. Anchor | B. Noise reduction | C. Proposal | B gain = B - A | C gain = C - A |
|-------|----------|-----------|--------------------|-------------|----------------|----------------|
| A | S1.  Traffic | 7.63 | 9.43 | 9.58 | 1.8 | 1.95 |
| A | S2.  PeopleOnStreet | 5.41 | 7.49 | 7.98 | 2.08 | 2.57 |
| A | S3.  SteamLocomotive | 8.15 | 9.31 | 9.41 | 1.16 | 1.26 |
|   | **Average – class A** | **7.06** | **8.74** | **8.99** | **1.68** | **1.93** |
| B1 | S4.  Kimono1 | 7.02 | 8.61 | 8.83 | 1.61 | 1.83 |
| B1 | S5.  ParkScene | 6.45 | 8.12 | 8.23 | 1.67 | 1.78 |
| B2 | S6.  Cactus | 7.52 | 8.62 | 8.87 | 1.19 | 1.35 |
| B2 | S7.  BQTerrace | 7.69 | 9.13 | 9.38 | 1.44 | 1.69 |
| B2 | S8.  BasketballDrive | 7.59 | 8.93 | 9.11 | 1.34 | 1.52 |
|   | **Average – class B** | **7.25** | **8.68** | **8.88** | **1.43** | **1.63** |
| C | S9.  RaceHorses | 5.36 | 6.88 | 7.2 | 1.52 | 1.84 |
| C | S10 .BQMall | 5.99 | 7.13 | 7.43 | 1.14 | 1.44 |
| C | S11. PartyScene | 4.69 | 6.17 | 6.61 | 1.47 | 1.91 |
| C | S12. BasketballDrill | 4.46 | 5.05 | 5.41 | 0.59 | 0.95 |
|   | **Average – class C** | **5.13** | **6.31** | **6.66** | **1.18** | **1.54** |
| D | S13. RaceHorsesLow | 5.25 | 7.01 | 7.25 | 1.76 | 2.01 |
| D | S14. BQSquare | 5.01 | 6.87 | 7.12 | 1.87 | 2.12 |
| D | S15. BlowingBubbles | 5.73 | 7.4 | 7.71 | 1.67 | 1.98 |
| D | S16. BasketballPass | 4.52 | 5.18 | 5.50 | 0.68 | 1.00 |
|   | **Average – class D** | **5.12** | **6.62** | **6.90** | **1.50** | **1.78** |
| E | S17. FourPeople | 5.31 | 6.31 | 6.51 | 1.01 | 1.19 |
| E | S18. Johnny | 4.37 | 6.73 | 6.89 | 2.36 | 2.52 |
| E | S19. KristenAndSara | 6.08 | 7.21 | 7.56 | 1.13 | 1.48 |
|   | **Average – class E** | **5.25** | **6.75** | **6.98** | **1.50** | **1.73** |
|   | **Average – all** | **6.01** | **7.45** | **7.71** | **1.44** | **1.71** |

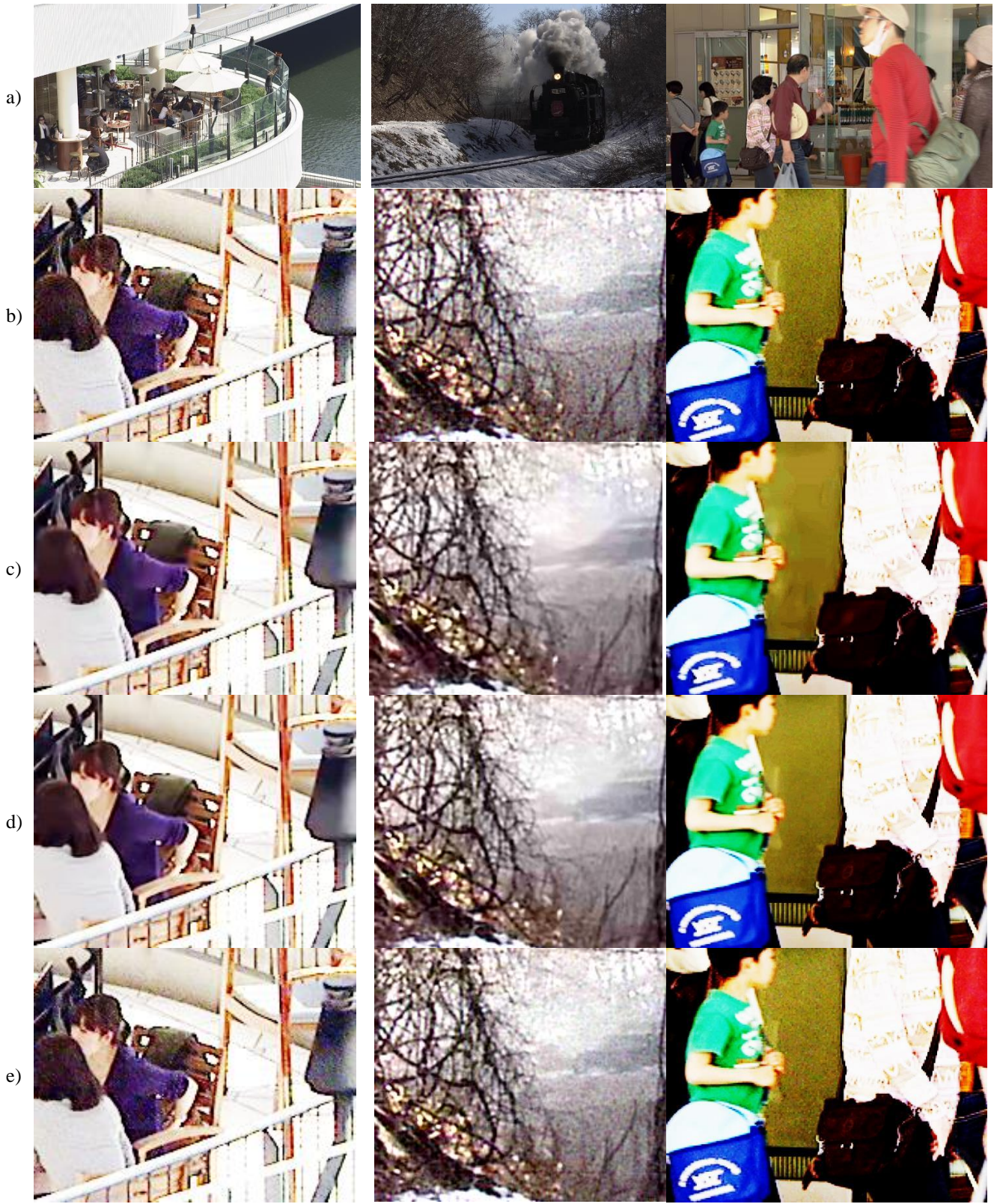

Fig. 10. Examples of images from sequences "BQterrace" (left), "SteamLocomotive" (middle) and "BQMall" (right).

a) Original frame, b) enlarged and contrast-enhanced original fragment.

Encoded videos (rate R4):   c) A.Anchor, d) B. Noise Reduction, e) C.Proposal.

Table 6. Execution time per single frame (average over sequences) of particular components of the considered video compression system.

| Component | Execution time [s] | | | | | | |
|---|---|---|---|---|---|---|---|
| | **Class A** | **Class B1** | **Class B2** | **Class C** | **Class D** | **Class E** | **Average** |
| Encoder | | | | | | | |
| Noise reduction | 2.52 | 4.12 | 1.18 | 0.24 | 0.05 | 0.46 | **1.43** |
| Base layer HEVC encoding | 260.36 | 432.61 | 130.26 | 28.43 | 6.25 | 51.56 | **151.58** |
| **Noise modeling and encoding** | 0.31 | 0.51 | 0.18 | 0.05 | 0.01 | 0.05 | **0.19** |
| Decoder | | | | | | | |
| Base layer HEVC decoding | 0.83 | 0.47 | 0.40 | 0.08 | 0.02 | 0.15 | **0.33** |
| **Noise layer decoding and synthesis** | 0.03 | 0.01 | 0.01 | 0.01 | 0.01 | 0.01 | **0.01** |

The Table 7 concludes the amount of data required for transmission of the proposed parametric model of the noise used in the experiments. For Spatial Density of Energy (SDE) constant QP=30 value has been used in HEVC encoding. In the case of Spectral Envelope (SE), as mentioned prediction of order $p = 10$ was used for both horizontal and vertical IIR filter parameters. This resulted in total 20 coefficients, each represented 8-bit LAR coefficients. Thus, for each frame, transmitted were 160 bits. Depending on the frame rate of particular sequences, this gives the bitrate required for representation of SE. Therefore depending on the rate-point and sequence (Table 1), the bitrate required for transmission of the proposed parametric model is about 0.2-5% of the total bitrate (Table 2). As it can be seen in Fig. 10, this is sufficient to represent a noise which is visually resembling the original noise. Also objectively (Fig. 11), the spectrum approximated with LPC method is in agreement with the spectrum of the original noise.

Table 7. The bitrate required for transmission of the proposed parametric model.

| Sequence class | SDE image size / number of HEVC Coding Units (CU) | Sequence | Frame rate | SE bitrate [kbit/s] | SDE bitrate [kbit/s] | Total noise model bitrate [kbit/s] |
|---|---|---|---|---|---|---|
| Class A | 85×53 / 2 | S1. Traffic | 30 | 4.8 | 7.4 | 11.8 |
| | | S2. PeopleOnStreet | 30 | 4.8 | 15.3 | 32.8 |
| | | S3. SteamLocomotive | 60 | 9.6 | 7.2 | 16.6 |
| Class B1 | 64×36 / 1 | S4. Kimono | 24 | 3.8 | 3.1 | 7.0 |
| | | S5. ParkScene | 24 | 3.8 | 4.9 | 8.7 |
| Class B2 | 64×36 / 1 | S6. Cactus | 50 | 8.0 | 8.8 | 16.8 |
| | | S7. BQTerrace | 60 | 9.6 | 8.8 | 18.4 |
| | | S8. Basketball Drive | 50 | 8.0 | 8.8 | 16.8 |
| Class C | 27×16 / 1 | S9. RaceHorses | 30 | 4.8 | 11.5 | 16.3 |
| | | S10. BQMall | 60 | 9.6 | 11.5 | 21.1 |
| | | S11. PartyScene | 50 | 8.0 | 19.2 | 27.2 |
| | | S12. BasketballDrill | 50 | 8.0 | 11.5 | 19.5 |
| Class D | 13×8 / 1 | S13. RaceHorsesLow | 30 | 4.8 | 13.7 | 18.5 |
| | | S14. BQSquare | 60 | 9.6 | 18.3 | 27.9 |
| | | S15. BlowingBubbles | 50 | 8.0 | 18.3 | 26.3 |
| | | S16. BasketballPass | 50 | 8.0 | 18.3 | 26.3 |
| Class E | 42×24 / 1 | S17. FourPeople | 60 | 9.6 | 3.5 | 13.1 |
| | | S18. Johnny | 60 | 9.6 | 2.1 | 11.7 |
| | | S19. KristenAndSara | 60 | 9.6 | 2.1 | 11.7 |

The bitrate reductions that can be attained with the use of the proposed method are not discussed in the present paper due to the following reasons:

- It is irrelevant to use Bjøntegaard [31] method based on PSNR, due to the addition of a synthetic noise to the decoded video, proposed in the present paper.

- The bitrate reduction originating from the use of sole noise reduction (case B), in a very similar coding scenario, has already been analyzed in the previous works [15], [16] and are reported to be about 50% on average.

- The bitrate of the proposed method (C) is practically the same as of sole noise reduction (B), because the bitstream related to the proposed noise model is negligible, of about 1%, depending on the case (see: Table 7), of the total bitrate.

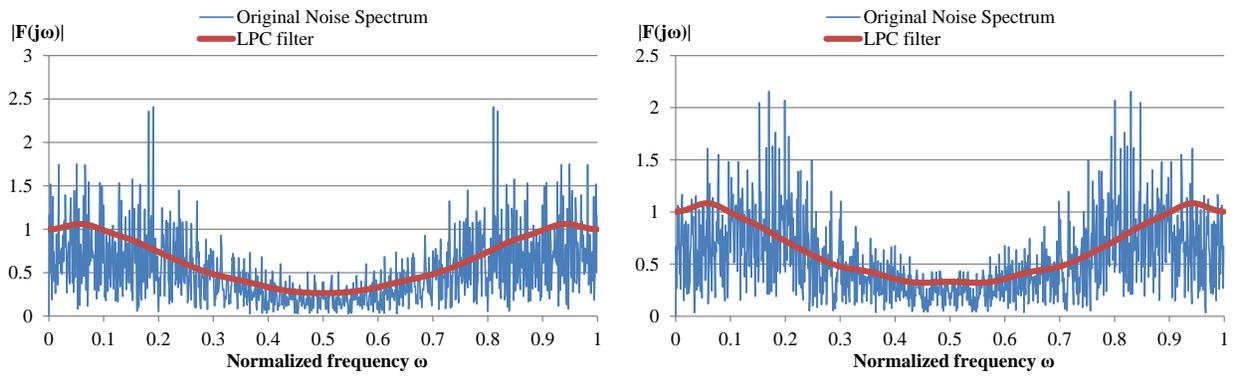

Fig. 10. Example of noise spectrum and its envelope approximation with the use of LPC filter in horizontal direction (left) and in vertical direction (right) for SteamLocomotion sequence.

## 5. Conclusions

A video encoding method in which noise is encoded using a novel parametric model has been proposed, and its implementation has been deliberated. The proposed model of noise consists of two components: Spatial Distribution of Energy (low-resolution video) and Spectral Envelope (encoded with LAR coefficients).

Although the proposal has been presented in the context of HEVC video coding technology, its idea is general and can be applied in conjunction with any general-purpose video codec.

Exhaustive subjective tests were performed to assess the performance of the proposed technique and to compare it with classical video coding and simpler methods which employ only noise reduction without noise modeling. The experiments confirm some results already attained in other works demonstrating that applying noise reduction prior to video compression result in considerable video quality increases of approximately 1.44 MOS points. The results also highlighted that the proposed model can be used to increase the quality of compressed video even further - by approximately 0.3-0.4 Mean Opinion Score (MOS) points. In total, the gain of the proposed method is about 1.71 MOS points, as compared to the classical video coding without any noise-related processing. Finally, it has been shown that use of the proposal is justified due to its low computational complexity.

Apart from interesting experimental results for this particular proposed and implemented technique, the very important outcome of the paper is evidence that the use of synthetic noise added to the video reconstructed at the decoder side can increase its subjective quality.

Avenues for the future research include using more advanced noise reduction techniques and investigating the proposed method in the context of different video codecs: monoscopic (e.g. AV1) or multiview (e.g. HEVC-3D).

### Acknowledgements

This work was supported by the Ministry of Science and Higher Education of Poland.